# New Light on Like-Charge Attraction

Jörg Baumgartl[1], Jose Luis Arauz-Lara[2], and Clemens Bechinger[1]

[1]2. Physikalisches Institut, Pfaffenwaldring 57, 70550 Stuttgart, Germany

[2] Instituto de Física, Universidad Autónoma de San Luis Potosí, Alvaro Obregón 64, 78000 San Luis Potosí, Mexico

We report on pair interaction measurements in charged colloidal systems in confined and unconfined geometries by means of optical microscopy. At very small particle distances we observe minute distortions in the optical images which can lead to artifacts in the pair potentials as derived by digital video microscopy. In particular, these distortions can pretend long-range attractions observed in charged colloidal suspensions.

A controversial debate in colloidal science has been launched in 1990 when Kepler and Fraden reported an unusual long-range attractive component in the pair potential of charged colloidal particles [1]. This so-called *like-charge attraction* (LCA) was only observed in thin sample cells (typical plate separations < 10µm) while the pair-interaction in unconfined suspensions has been experimentally confirmed to be entirely repulsive which is in agreement with Poisson-Boltzmann theory [2-4]. Therefore, it was speculated that the confining plates are responsible for the deviations from theory. Soon after its initial observation LCA was also observed by other authors [5-7] which then provoked considerable theoretical interest in this phenomenon. In the meantime it has been rigorously proven that the observed attraction can not be explained within the framework of mean field theories, irrespective of whether the particles are suspended in bulk or in confinement [8-10]. Several other approaches beyond Poisson-Boltzmann were proposed as the origin of confinement-induced attraction. However, they are controversially discussed [11,12] and seem to fail to reproduce the experimental observations thus making LCA a persistent mystery.

In this Letter we reinvestigate the pair-potential of charged colloidal particles in confined and unconfined geometries. In contrast to previous experiments where pair potentials U(r) were derived from pair-correlation functions g(r), here we measure U(r) directly between two silica spheres confined in an extended optical trap. This method has been successfully applied by several groups to study other interaction types in colloidal suspensions [4,13]. We demonstrate that optical artifacts caused by the imaging process can lead to minute distortions in the particle distances as obtained by digital video microscopy. Those distortions result in an apparent minimum in U(r) which agrees with respect to its position and depth with the features observed in LCA. After correction of these distortions we obtain - independent of the confinement conditions - entirely repulsive pair interactions which show good agreement with linearized mean field theories.

Interaction potentials between colloids were obtained by subjecting two particles to a well-defined radially symmetric light potential which was created by a slightly defocused Gaussian laser beam ($TEM_{00}$, $\lambda = 532$ nm). The light pressure from the vertically incident laser beam confines the particles irrespective of the cell height to two dimensions [14]. From the particle trajectories as measured by digital video microscopy (for a review see e.g. [15]) we obtain their relative distance distribution P(r) which then leads to the pair-potential $P(r)=P_0 \exp(-\{U_{ext}+U(r)\}/k_B T)$ with $U_{ext}$ the potential induced by the laser tweezers. In the central region which is sampled by the particles we confirmed that $U_{ext}$ is approximated very well by a parabolic potential. This shape is particularly convenient because in parabolic potentials the center of mass motion decouples from the relative particle motion and thus allows to obtain $U_{ext}$ and U(r) simultaneously (otherwise the external potential must be first determined by the probability distribution of a single colloidal particle inside the laser trap [4]). All pair potentials as shown in the following were reproduced for different laser intensities up to about 50mW resulting in different optical trapping strengths. This was done to rule out possible light-induced effects such as optical binding [16].

To investigate possible confinement-induced effects, we studied U(r) in aqueous suspensions of charged silica particles of diameter $\sigma = 1.5 \pm 0.08$ µm in different cell geometries. For unconfined conditions we used commercially available glass cuvettes with 200µm spacing which were coupled to a deionizing circuit to maintain stable ionic conditions at different ionic strengths. Thin cells were fabricated according to a procedure as described in [17]. Briefly, bidisperse suspensions containing the above silica particles and a small amount of slightly larger polystyrene particles ($\sigma = 1.96$ µm) were dialyzed against pure water for several weeks to obtain good deionization. A small amount of this mixture was confined between two clean glass plates which were uniformly pressed on top of each other until the separation between the glass plates is determined by the larger particles. Accordingly, the larger particles which serve as spacers for the two plates were randomly distributed across the sample while the smaller ones are still mobile. Afterwards the system was sealed with epoxy resin which yielded stable conditions over several weeks. Since the difference in diameter between the large and the small spheres is rather small, vertical fluctuations of the small particles are largely reduced. In previous experiments it has been demonstrated that the pair potential of the smaller particles in such cells obtained via measuring g(r) shows a long-range attraction of several tenths of $k_BT$ at $r \approx 1.5$ $\sigma$ [5,7].

Fig.1 shows measured pair potentials between two silica spheres inside the 200µm thick cuvette under different conditions as described below. Fig.1A corresponds to the pair potential for an ionic conductivity of about 0.8 µS/cm. The solid line is a fit to a screened

Coulomb potential $U(r) = (Z^*)^2 \lambda_B \left( \frac{\exp(\kappa R)}{1 + \kappa R} \right)^2 \frac{\exp(-\kappa r)}{r}$ with $Z^*$ the effective colloidal charge, R the particle radius, $\kappa$ the inverse Debye screening length, and $\lambda_B$ = 0.72nm the Bjerrum length in water. We yield $Z^*$ = 15900 ± 1100 and $\kappa^{-1} \approx$ 190nm, the latter being in the range of the value expected according to the micromolecular strength in the sample cell. Fig.1B shows the pair potential after increasing the ionic conductivity to 4 µS/cm. As expected, the potential is shifted by about 500nm to the left thus indicating the higher screening in the suspension. In addition, however, a pronounced minimum is observed. This minimum is with respect to its position ($\approx 1.5\sigma$) and depth ($\approx 0.3 k_BT$) in excellent agreement with the characteristics of LCA in confined systems [1,5,7]. Interestingly, such a minimum is observed here in a thick cell which demonstrates that attractive parts in the pair potential may be also observed in the absence of confinement. Fig.1C shows the result obtained under the identical sample conditions as in Fig.1B but now under slightly different optical illumination conditions (see insets in Fig.1). While in Figs.1A,B the particles were illuminated from above with convergent white light (using a condenser lens), Fig.1C was obtained for quasi-parallel illumination. The numerical aperture (NA) of the microscope objective used for particle imaging onto a CCD camera was always identical, i.e. NA=0.75. In addition to the minimum at r $\approx 1.5\sigma$, Fig.1C displays an oscillatory behavior with a pronounced second minimum at about 2.5$\sigma$. Obviously, the illumination conditions strongly affect the measured pair potentials and it is helpful to look at typical snapshots of the colloidal pairs obtained under convergent and quasi-parallel illumination (upper and lower inset of Fig.1). While in the first case the particles are imaged as bright central spots with a dark ring, in the second case two additional concentric bright rings are visible. To understand the appearance of the ring system around the particles we want to recall that imaging processes correspond to a double Fourier transform and according to Abbe it is the highest transmitted diffraction order which eventually limits the optical resolution. Consequently, a perfect image is only obtained if *all* diffraction orders contribute to the image. Since microscopes have a finite NA, they do not transmit all diffraction orders (spatial filtering) and distortions as seen in the insets of Fig.1 are unavoidable. The extent of these artifacts critically depend on the illumination conditions, the NA used for optical imaging but also the particle size and their index of refraction (see e.g. [18]). As we will demonstrate in the following these ring systems are responsible for slight distortions in the imaging process which then lead to slight artifacts in the particle center positions as determined by digital video microscopy.

For quantitative investigations of how the optical artifacts affect measured particle distances we performed experiments under the same conditions as in Fig.1B but with one of the particles tightly bound to the substrate to act as an immobile positional reference point. In our sample cells we always observed some of the particles irreversibly adsorbed to the walls due to van-der Waals forces. From the center of the probability distribution of the reference particle (our positional resolution is about 25nm) we obtained its true position $(\overline{x}_{ref}, \overline{y}_{ref})$. Next, we inserted a free fluctuating particle into the laser trap which was about 10σ apart from the immobile one. At those distances no optical artifacts occur. From the particle trajectories we now calculate $\Delta r(t) = r(t) - \sqrt{[x(t) - \overline{x}_{ref}]^2 + [y(t) - \overline{y}_{ref}]^2}$ with $r(t)$ the measured particle distance and $(x(t), y(t))$ the position of the free fluctuating colloid as determined by video microscopy (Fig.2A,C). As can be seen the data points scatter within our experimental resolution around $\Delta r = 0$. When we repeat this experiment with the free particle close to the adsorbed one, the fluctuations in $\Delta r$ seem to be dramatically increased and the histogram becomes asymmetrically broadened (Fig.2B,D). This clearly demonstrates that in this case the measured particle distances have to be taken with care. In the inset of Fig.3 we plotted Δr vs. r which underlines that there is a characteristic distance dependence on the aforementioned optical distortion. Owing to the symmetry of the problem, the distortion affects both - the adsorbed and the free- particles in the same way, therefore the true particle distance $r_t$ is simply given by $r_t = r - 2\Delta r$. Since $\Delta r$ changes its sign at small r, this means that for $r < 1.7\sigma$ the particles are closer than they appear while the opposite is found for $1.7\sigma < r < 2.6\sigma$. Using the smoothed $\Delta r$ vs. r curve (black line) we can now correct the probability distribution P(r) as obtained in the previous experiments and replot them as $P(r_t)$ which then yields the corrected pair potential. Fig.3 shows the data of Fig.1B before and after the correction as open and closed symbols, respectively. After the correction the minimum disappeared and the pair-potential agrees well with a screened Coulomb potential yielding $Z^* = 18700 \pm 1500$ and $\kappa^{-1} = 55$ nm (solid line). This screening length is in very good agreement with the corresponding value derived from the ionic conductivity, i.e. $\kappa^{-1} \approx 50$ nm.

Similar measurements with different illumination conditions and microscope objectives were also performed in confined systems. Here we show the results obtained with an oil immersion objective (100x, NA=1.25) which has been also used by other authors. The corresponding pair potential obtained from the uncorrected data is shown as open symbols in Fig.4 and again displays an attractive component below $1.3\sigma$ with a depth of about $0.2 k_BT$. This is in agreement with LCA observed previously by other authors in confined geometries [1,19]. Due to the large NA used in this experiment, $\Delta r$ is less pronounced compared to Fig.3 but looks qualitatively very similar (see lower inset of Fig.4). Correcting finally the measured pair potential and replotting it versus the true particle distance leads again to the disappearance of the minimum as observed before. Even more interestingly is the fact, that the resulting U(r) under confinement can be well fitted to a Yukawa potential (solid line) with $Z^* \approx 8000 \pm 4000$ and $\kappa^{-1} \approx 10 \pm 5$ nm.

Our results also provide a simple explanation why LCA was only reported in confined geometries and no attractive components have been observed in unconfined systems. Due to the counter ions of the charged glass plates the effective screening length is rather small in thin cells. As demonstrated, however, small distances are particularly susceptible to the observed distortion effects. In contrast, interaction potential studies in unconfined geometries have been typically performed at smaller microionic concentrations where particle distances below 2σ are hardly sampled [2-4,20]. In this case optical distortion effects can be neglected and digital video microscopy yields accurate results. Our findings are also consistent with an (so far unexplained) observation made in the first paper on LCA where it was demonstrated that even under confinement, attractive contributions in U(r) only occur at sufficiently high ionic concentrations, i.e. at sufficiently close particle distances [1].

In conclusion, we can not support attractive components in the pair-interaction of confined colloidal suspensions but rather suggest that LCA is caused by optical artifacts resulting from spatial filtering during the imaging process. Therefore, special care has to be taken in the interpretation of small particle distances as obtained by video microscopy.

We want to thank A. Ramirez-Saito for preparation of thin sample cells and L. Helden, V. Blickle and D. Babic for fruitful discussions.

**Figure Captions**

**Fig.1** Pair potentials of colloidal particles in a 200μm thick cell obtained by video microscopy: (A) illuminated with white convergent light at moderate ionic strengths. (B) same illumination conditions as A but higher salt concentrations. (C) same salt conditions as B but illumination with quasi-parallel light. The insets show typical snapshots for the convergent (upper) and quasi-parallel (lower picture) illumination conditions.

**Fig.2** $\Delta r(t)$ between a fixed and free particle (A) for large and (B) for small particle distances. The data in B are shifted for clarity by 0.05σ in vertical direction. (C,D) Corresponding histograms of the measured distance fluctuations.

**Fig.3** Corrected (closed symbols) and vertically shifted uncorrected (open symbols) colloidal pair potentials obtained in a 200µm thick cell. Inset: Δr obtained from Fig.2 vs. r (grey). The solid line is obtained after smoothing the data.

**Fig.4** Corrected (closed symbols) and vertically shifted uncorrected (open symbols) pair potential in confined geometry (cell thickness 1.96µm). The upper and lower insets show a snapshot and the Δr vs. r curve.

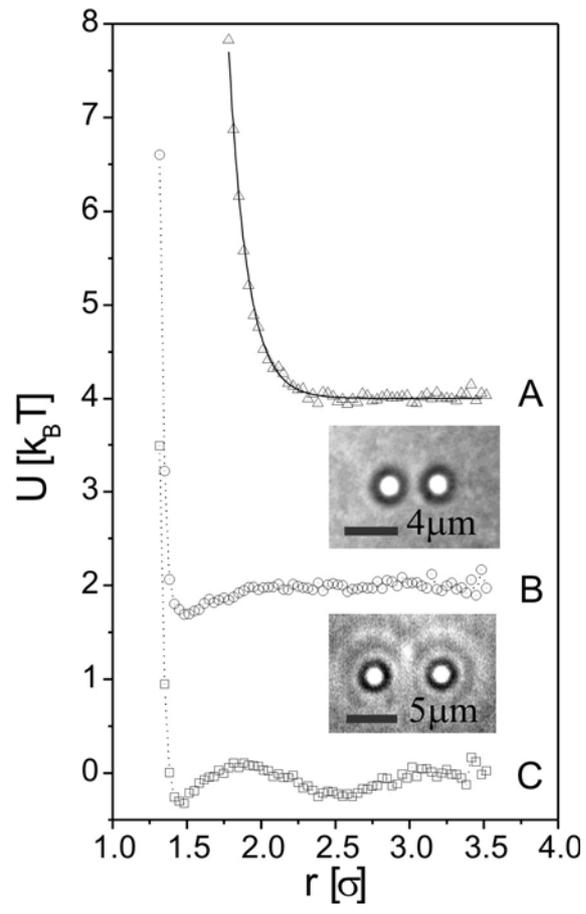

Fig.1

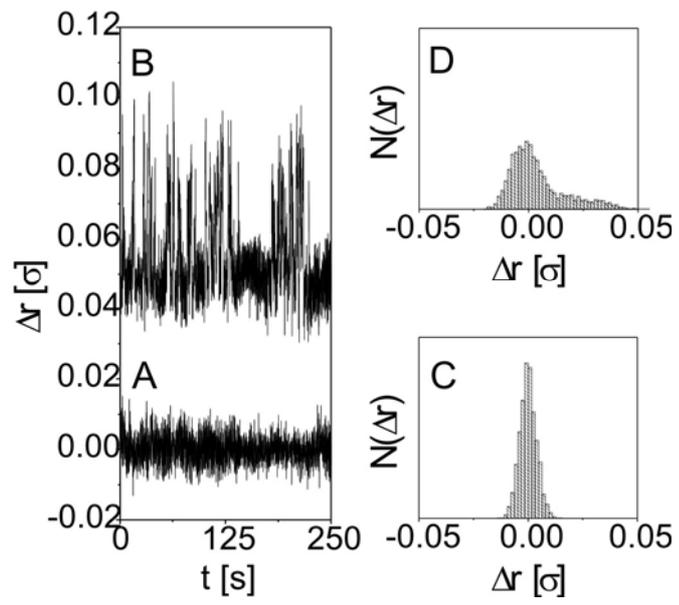

Fig.2

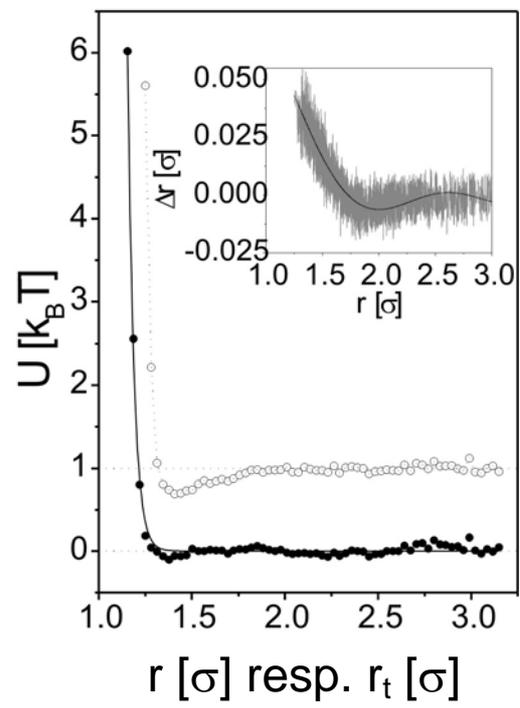

Fig.3

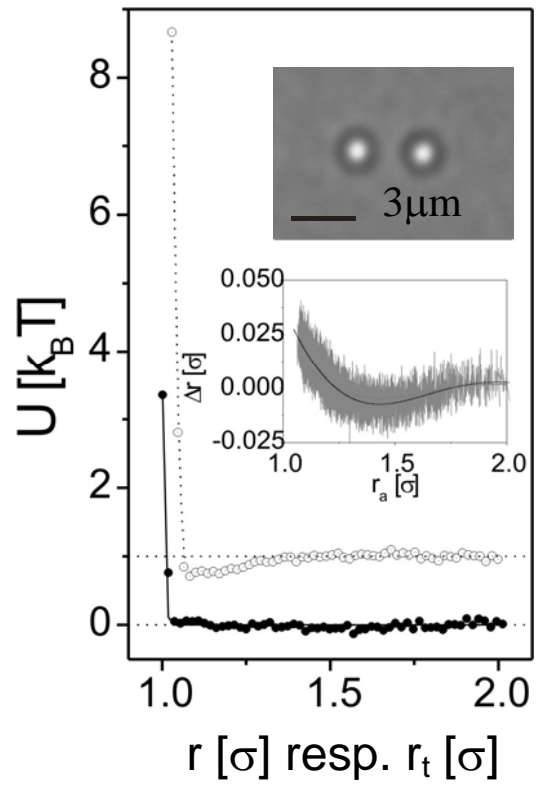

Fig.4